\documentclass[aps,showpacs,nofootinbib,superscriptaddress,10pt]{revtex4}
\usepackage{graphics}
\usepackage{graphicx}
\usepackage{amssymb,amsmath}
\RequirePackage{xspace}

\def\Kbar  {\kern 0.2em\overline{\kern -0.2em K}{}\xspace}

\begin{document}

\title{Measurement of the $\mathbf{e^+e^- \rightarrow K_S K_L \pi^0}$
cross section in the energy range $\mathbf{\sqrt{s}=1.3-2.0}$ GeV}

\begin{abstract}
The $e^+e^- \rightarrow K_S K_L \pi^0$ cross section is measured in the 
center-of-mass energy range $\sqrt{s}=1.3-2.0$ GeV. The analysis is based on 
the data sample with an integrated luminosity of 33.5 pb$^{-1}$
collected with the SND detector at the VEPP-2000 $e^+e^-$ collider.
\end{abstract}

\author{M.~N.~Achasov}
\author{V.~M.~Aulchenko}
\author{A.~Yu.~Barnyakov}
\author{K.~I.~Beloborodov}
\author{A.~V.~Berdyugin}
\author{D.~E.~Berkaev}
\affiliation{Budker Institute of Nuclear Physics, SB RAS, Novosibirsk, 630090,
Russia}
\affiliation{Novosibirsk State University, Novosibirsk, 630090, Russia}
\author{A.~G.~Bogdanchikov}
\author{A.~A.~Botov}
\affiliation{Budker Institute of Nuclear Physics, SB RAS, Novosibirsk, 630090,
Russia}
\author{T.~V.~Dimova}
\author{V.~P.~Druzhinin}
\author{V.~B.~Golubev}
\author{L.~V.~Kardapoltsev}
\affiliation{Budker Institute of Nuclear Physics, SB RAS, Novosibirsk, 630090,
Russia}
\affiliation{Novosibirsk State University, Novosibirsk, 630090, Russia}
\author{A.~S.~Kasaev}
\affiliation{Budker Institute of Nuclear Physics, SB RAS, Novosibirsk, 630090,
Russia}
\author{A.~G.~Kharlamov}
\affiliation{Budker Institute of Nuclear Physics, SB RAS, Novosibirsk, 630090,
Russia}
\affiliation{Novosibirsk State University, Novosibirsk, 630090, Russia}
\author{A.~N.~Kirpotin}
\affiliation{Budker Institute of Nuclear Physics, SB RAS, Novosibirsk, 630090,
Russia}
\author{I.~A.~Koop}
\affiliation{Budker Institute of Nuclear Physics, SB RAS, Novosibirsk, 630090,
Russia}
\affiliation{Novosibirsk State University, Novosibirsk, 630090, Russia}
\affiliation{Novosibirsk State Technical University, Novosibirsk, 630092, Russia}
\author{L.~A.~Korneev}
\email[e-mail:]{leonidkorneev@gmail.com}
\affiliation{Budker Institute of Nuclear Physics, SB RAS, Novosibirsk, 630090,
Russia}
\author{A.~A.~Korol}
\affiliation{Budker Institute of Nuclear Physics, SB RAS, Novosibirsk, 630090,
Russia}
\affiliation{Novosibirsk State University, Novosibirsk, 630090, Russia}
\author{D.~P.~Kovrizhin}
\author{S.~V.~Koshuba}
\affiliation{Budker Institute of Nuclear Physics, SB RAS, Novosibirsk, 630090,
Russia}
\author{A.~S.~Kupich}
\author{N.~A.~Melnikova}
\affiliation{Budker Institute of Nuclear Physics, SB RAS, Novosibirsk, 630090,
Russia}
\affiliation{Novosibirsk State University, Novosibirsk, 630090, Russia}
\author{K.~A.~Martin}
\author{A.~E.~Obrazovsky}
\author{A.~V.~Otboev}
\author{E.~V.~Pakhtusova}
\affiliation{Budker Institute of Nuclear Physics, SB RAS, Novosibirsk, 630090,
Russia}
\author{K.~V.~Pugachev}
\author{Yu.~A.~Rogovsky}
\author{A.~I.~Senchenko}
\author{S.~I.~Serednyakov}
\author{Z.~K.~Silagadze}
\author{Yu.~M.~Shatunov}
\affiliation{Budker Institute of Nuclear Physics, SB RAS, Novosibirsk, 630090,
Russia}
\affiliation{Novosibirsk State University, Novosibirsk, 630090, Russia}
\author{D.~A.~Shtol}
\affiliation{Budker Institute of Nuclear Physics, SB RAS, Novosibirsk, 630090,
Russia}
\author{D.~B.~Shwartz}
\affiliation{Budker Institute of Nuclear Physics, SB RAS, Novosibirsk, 630090,
Russia}
\affiliation{Novosibirsk State University, Novosibirsk, 630090, Russia}
\author{I.~K.~Surin}
\author{ Yu.~V.~Usov}
\affiliation{Budker Institute of Nuclear Physics, SB RAS, Novosibirsk, 630090,
Russia}
\author{A.~V.~Vasiljev}
\affiliation{Budker Institute of Nuclear Physics, SB RAS, Novosibirsk, 630090,
Russia}
\affiliation{Novosibirsk State University, Novosibirsk, 630090, Russia}

\maketitle

\section{Introduction}
This paper is dedicated to the study of the reaction 
$e^+e^- \rightarrow K_S K_L \pi^0$.   
This reaction is one of three charge modes of the process 
$e^+e^- \rightarrow K\Kbar\pi$, which gives a sizable contribution
to the total cross section of $e^+e^-$ annihilation into hadrons in
the center-of-mass energy range $\sqrt{s} = 1.5-1.8$ GeV. 
The process $e^+e^- \rightarrow K\Kbar\pi$ is also important for
spectroscopy of $s\bar{s}$ vector states. From these states, only the
lowest $\phi(1020)$ is well studied. In particular, its branching fractions
are measured up to $10^{-5}$ level. Spectroscopy of the first excited state 
$\phi(1680)$ is far from completion. The main decay mode of
$\phi^\prime\equiv\phi(1680)$ is $K^\ast(892) \Kbar$\footnote{Throughout 
this paper the use of charge conjugate modes is implied.}
with the $K^\ast(892)$ decay to $K\pi$.

Processes of $e^+e^-$ annihilation to the $K \Kbar \pi$ final state  were 
studied in the DM1, DM2 and BABAR~\cite{dm1,dm2,babar_kkp,babar}
experiments. Until recently, only the two subprocesses $e^+e^- \rightarrow
K_S K^{\pm} \pi^{\mp}$, $K^+ K^{-} \pi^{0}$~\cite{babar_kkp} were measured.
The third, neutral subprocess $e^+e^- \rightarrow K_S K_L \pi^0$,
is hard to study due to complexity of $K_L$-meson detection and 
identification.  Recently, it was measured in the BABAR
experiment~\cite{babar}. The measurement uses the initial state radiation 
method, in which the $e^+e^- \rightarrow X$ cross section is determined from
the mass spectrum of the hadron system $X$ in the reaction 
$e^+e^- \rightarrow X\gamma$. Detection of all final 
particles in the reaction $e^+e^- \rightarrow K_S K_L \pi^0\gamma$ was
required, and the $K_L$ meson was identified as a single photon. 
The efficiency of $K_L$-meson detection was measured using   
$e^+e^- \to \phi(1020)\gamma\to K_S K_L\gamma$ events selected 
without any conditions on $K_L$ parameters.
It should be noted that the $K_L$-meson energy was not measured. Therefore,
good background suppression was not reached in Ref.~\cite{babar}.
A relatively large systematic uncertainty of the 
$e^+e^- \rightarrow K_S K_L \pi^0$ cross-section measurement ($\sim 10\%$
at the cross-section maximum) is due to an uncertainty in background 
subtraction.

In this paper, the process $e^+e^- \rightarrow K_{S} K_{L} \pi^0$ is studied
using a data sample collected in the energy range $\sqrt{s}=1.3-2.0$ GeV 
with the SND detector~\cite{SND} at the VEPP-2000 $e^+e^-$ 
collider~\cite{VEPP2000}.

\section{Detector and experiment}
SND is a general purpose nonmagnetic detector. Its main part  
is a three-layer electromagnetic calorimeter based on NaI(Tl) crystals.
The calorimeter covers a solid angle of 95\% of $4\pi$. Its energy resolution
is $\sigma_{E_\gamma}/E_\gamma=4.2\%/\sqrt[4]{E_\gamma(\mbox{GeV})}$,
while the angular resolution is 
$\sigma_{\theta,\phi}=0.82^{\circ}/\sqrt{E_\gamma(\mbox{GeV})}$,
where $E_\gamma$ is the photon energy.
The tracking system is located inside the calorimeter, around the collider 
beam pipe. It consists of a nine-layer cylindrical drift chamber and a 
proportional chamber with cathode-strip readout. A solid angle covered 
by the tracking system is 94\% of $4\pi$. For charged kaon identification, 
the system of threshold aerogel Cherenkov counters is used.
The calorimeter is surrounded by a muon system
consisting of proportional tubes and scintillation counters.

The analysis uses a data sample with an integrated luminosity of
33.5 pb$^{-1}$ recorded in 2010--2012 in the energy region 1.3--2.0 GeV.
Due to relatively small statistics of selected events of the process under
study, data collected in 36 energy points are combined into 15 energy
intervals listed in Table~\ref{table1}.

The simulation of the process $e^+e^- \rightarrow K_{S} K_{L} \pi^0$
is performed using a Monte-Carlo (MC) event generator based on formulas from 
Ref.~\cite{wppp}. It is assumed that the process proceeds via the
$K^\ast(892)^0\Kbar^0$ intermediate state.

Interaction of particles produced in $e^+e^-$ collision with
the detector material is simulated using GEANT4 v.9.5 package~\cite{geant4}.
Analyses of processes with $K_{L}$ meson in the final state 
critically depend on correct simulation of $K_{L}$ nuclear
interaction. Unfortunately, both the total and inelastic low-energy
cross section of the $K_{L}$ nuclear interaction are strongly overestimated 
in GEANT4 v.9.5~\cite{len}. Therefore, we have modified the GEANT4
module responsible for $K_{L}$ cross-section calculation using 
the model from Ref.~\cite{kloe}. This model describes reasonably well
the experimental data both on the total $K_{L}$ cross section
in different materials (H, Be, C, Al, Fe, Cu, Pb) in the $K_{L}$ energy range
525--600 MeV~\cite{totcs} and on the inelastic cross section
in the range 510--700 MeV~\cite{len}. Accuracy of the model
is estimated by comparison of its prediction with the precisely measured value
of the $K_{L}$ inelastic cross section at 510 MeV~\cite{len} and is found to 
be about 12\%.

The simulation takes into account variation of experimental conditions during
data taking, in particular dead detector channels, size and position of the
collider interaction region, beam-induced background etc. 
The beam background leads to appearance of spurious photons and/or 
charged particles in data events. To take this effect 
into account in simulation,  special background events are recorded during 
data taking with a random trigger. These events are then superimposed on 
simulated events.

In this paper the reaction $e^+e^- \rightarrow K_{S} K_{L} \pi^0$ is 
studied in the decay mode $K_{S} \to 2\pi^0$,
with no charged particles in the final state. Therefore, the process 
$e^+e^- \rightarrow \gamma \gamma$ is used for normalization. 
As a result of the normalization a part of systematic uncertainties
associated with event selection criteria for the process under study
is canceled out. The accuracy of the luminosity measurement using
$e^+e^- \rightarrow \gamma \gamma$ was studied in Ref.~\cite{ppg_snd} and 
is estimated to be 1.4\%.

\section{Event selection}
The reaction $e^+e^- \rightarrow K_{S} K_{L} \pi^0$ is studied 
in the $K_{S}  \rightarrow \pi^0\pi^0$ decay mode. The $K_{L}$ decay length
is much larger than the radius of the SND calorimeter, and the length of its
inelastic nuclear interaction is comparable with the calorimeter 
thickness~\cite{len}. In a significant fraction of 
$e^+e^- \rightarrow K_{S} K_{L} \pi^0$ events ($25-30\%$)
$K_{L}$ meson does not interact with the calorimeter, and
only six photons from decays of three $\pi^0$ are detected.
The $K_{L}$ meson undergoing a nuclear interaction inside the detector
produces one or several clusters in the calorimeter, which are
reconstructed as photons.

The selection of $e^+e^- \rightarrow K_{S} K_{L} \pi^0$ events is based
on finding three pairs of photons forming three $\pi^0$ candidates.
Two of these $\pi^0$'s having the invariant mass close to the $K^0$ mass
form a $K_{S}$ candidate. The events of the process under study are selected
in two stages. The primary selection is based on the following criteria:
\begin{itemize}
\item No charged tracks are reconstructed in the drift chamber.
The number of hits in the drift chamber is less than four.
\item The fired calorimeter crystals do not lie along a straight line.
This requirement rejects cosmic-ray background. 
\item An event contains at least six ``good'' photons ($N_\gamma\ge 6$) and no 
charged particles. A ``good'' photon is a cluster in the calorimeter with the
energy deposition larger than 20 MeV, which has a transverse energy profile
consistent with expectations for a photon~\cite{XINM}.
The latter condition rejects spurious photons originating from 
$K_L$ nuclear interaction or decay.
\item There are three $\pi^0$ candidates in an event. The $\pi^0$ candidate
is a pair of photons with the invariant mass in the range from 110 to 
160 MeV/$c^2$. 
\item The invariant mass of two $\pi^0$ candidates lies in the range
450--550 MeV/$c^2$.
\end{itemize}

The fraction of signal events rejected by the condition on the number of hits
in the drift chamber varies between 6\% and 17\%, depending on machine
background conditions. It should be noted that the same condition is 
used for selection of $e^+e^-\to \gamma\gamma$ events. Therefore, 
the possible systematic uncertainty due to this condition cancels as a result
of the luminosity normalization.

For energies above 1.9 GeV an additional selection criterion is applied to 
suppress the background from the process $e^+e^- \rightarrow K_{S}K_{L}\pi^0\pi^0$,
$K_{S} \rightarrow \pi^0\pi^0$. Events containing more than three $\pi^0$ 
candidates selected with the mass window 100--170 MeV/$c^2$ are rejected.  

The selected events are then kinematically fitted with three $\pi^0$ 
mass constraints and a $K_{S}$ mass constraint. The $\chi^2$ of the
kinematic fit is required to be less than 15. The refined photon
parameters are used to calculate the mass recoiling against the $K_{S}\pi^0$
system ($M_{\rm rec}$).

\section{Analysis of intermediate states in the reaction
$\mathbf{e^+e^- \rightarrow K_{S} K_{L} \pi^0}$\label{intst}}
In Ref.~\cite{babar} it is shown that the dominant mechanism of the 
$e^+e^- \rightarrow K_{S} K_{L} \pi^0$ reaction is the transition via
the $K^\ast(892)^0 \Kbar^0$ intermediate state. The fraction of 
$e^+e^- \to \phi \pi^0 \to K_{S} K_{L} \pi^0$ events near the maximum of the 
$e^+e^- \rightarrow K_{S} K_{L} \pi^0$ cross section (1.7 GeV) is about 
1\%~\cite{babar_kkp}. Also a small contribution of 
the $K_2^\ast(1430)^0 \Kbar^0$ state was observed in Ref.~\cite{babar}. 
The $e^+e^- \rightarrow K_2^\ast(1430)\Kbar$ cross section was measured in
Ref.~\cite{babar_kkp} in the charge modes $K^+K^-\pi^0$
and $K_S K^{\pm} \pi^{\mp}$. It proceeds in $D$ wave and is expected to be
negligibly small in the VEPP-2000 energy region, below 2 GeV.
\begin{figure}
\centering
\includegraphics[width=0.5\textwidth]{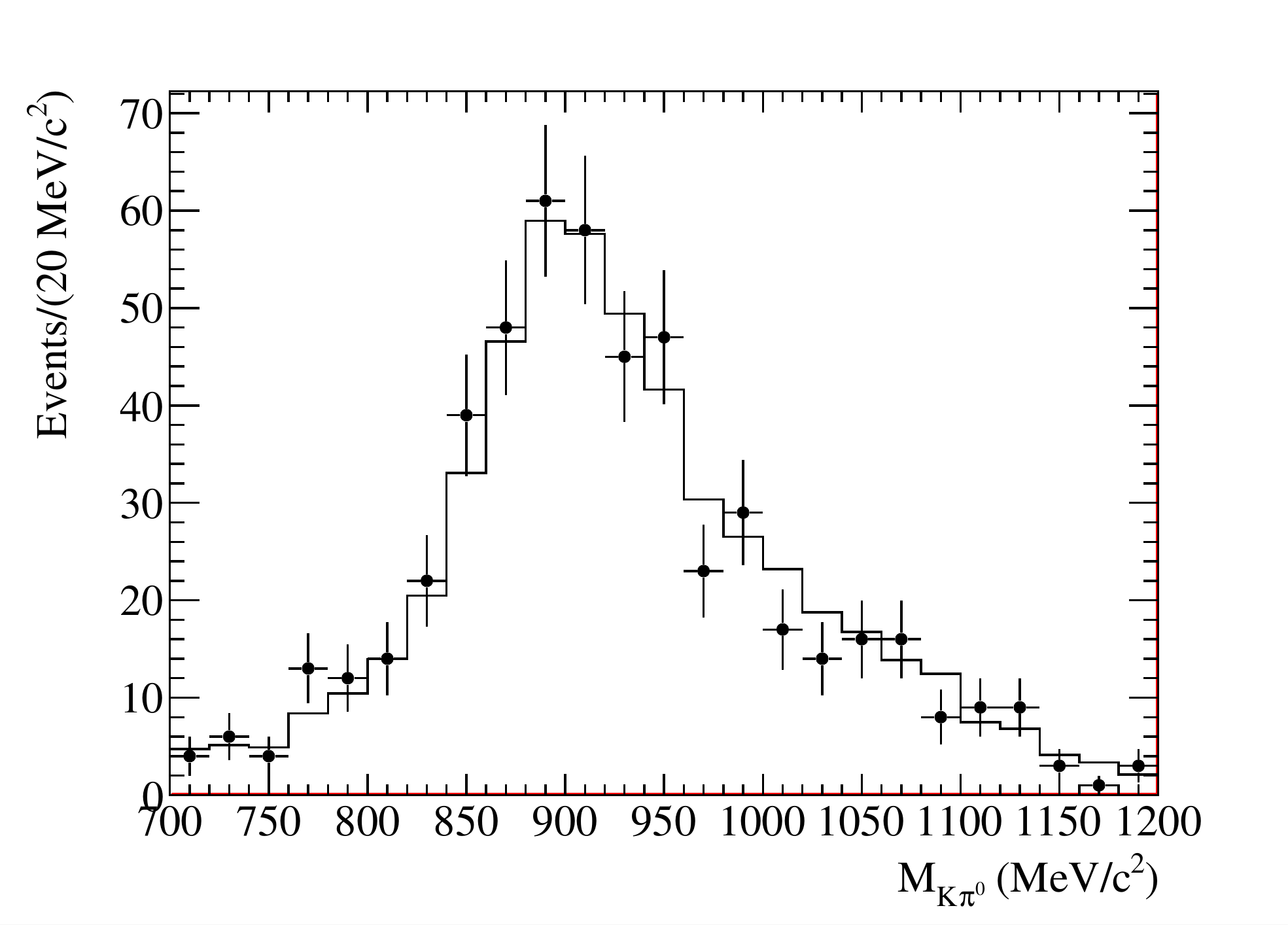}
\caption {The distribution of the invariant $K_{L} \pi^0$ and $K_{S} \pi^0$ 
masses (two entries per event) for data events from the energy region 
$\sqrt{s}=1.600-1.750$ GeV (points with error bars). The histogram
represents the simulated distributions obtained in the model with the
$K^\ast(892)^0 \Kbar^0$ intermediate state.
\label{fig_kpi}}
\end{figure}
\begin{figure}
\centering
\includegraphics[width=0.5\textwidth]{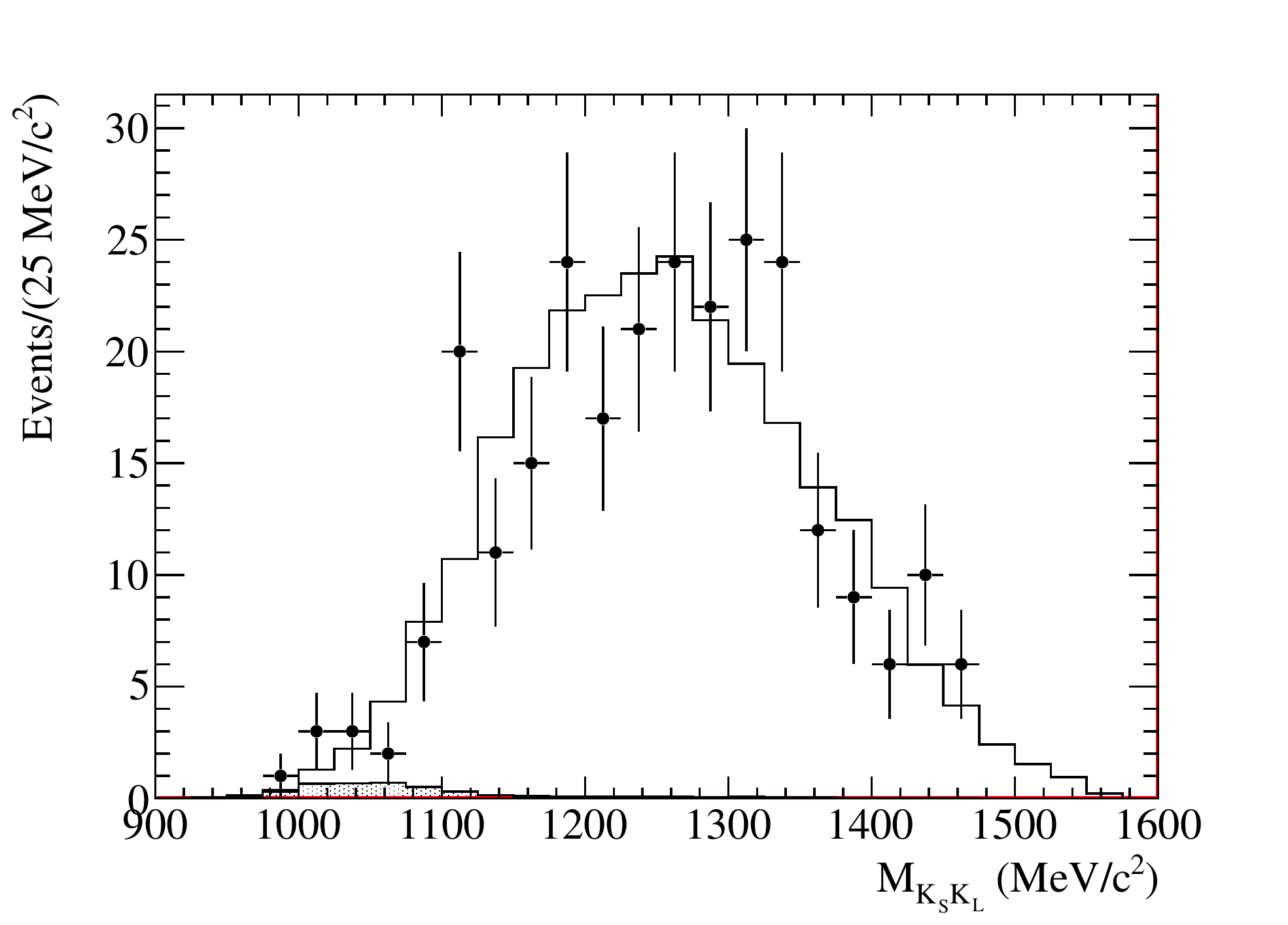}
\caption {The distribution of the $K_{S} K_{L}$ invariant
mass for data events from the energy region
$\sqrt{s}=1.600-1.750$ GeV (points with error bars). The histogram
represents the simulated distributions obtained in the model with the
$K^\ast(892)^0 \Kbar^0$ intermediate state. The shaded histogram
represents the expected contribution of the
$e^+e^-\to\phi\pi^0$ process estimated using MC simulation.
\label{fig_kk}}
\end{figure}

Figures~\ref{fig_kpi} and \ref{fig_kk} represent the distributions of the 
$K_{S(L)} \pi^0$ invariant mass (two entries per event) and
the $K_{S}K_{L}$ invariant mass, respectively, for six-photon data events 
from the energy region $\sqrt{s}=1.600-1.750$ GeV selected with the extra 
condition $400<M_{\rm rec}<600$ MeV/$c^2$. Six-photon events are used
to maximize the signal-to-background ratio. The fraction of background events 
in these distributions is estimated on the tails of the $M_{\rm rec}$ 
distribution (see Sec.~\ref{specfit}) and does not exceed 3\%. It is seen, 
that the data spectra in Figs.~\ref{fig_kpi} and \ref{fig_kk} are in good 
agreement with the simulated spectra obtained in the model with the 
$K^\ast(892)^0 \Kbar^0$ intermediate state. The shaded histogram
in Fig.~\ref{fig_kk} represents the expected contribution of the
$e^+e^-\to\phi\pi^0$ process estimated using MC simulation.  
With current statistics, we cannot observe the signal of the $\phi\pi^0$ 
intermediate state.

\section{Fit to the $\mathbf{M_{\rm rec}}$ spectrum\label{specfit}}
\begin{figure}
\centering
\includegraphics[width=0.5\textwidth]{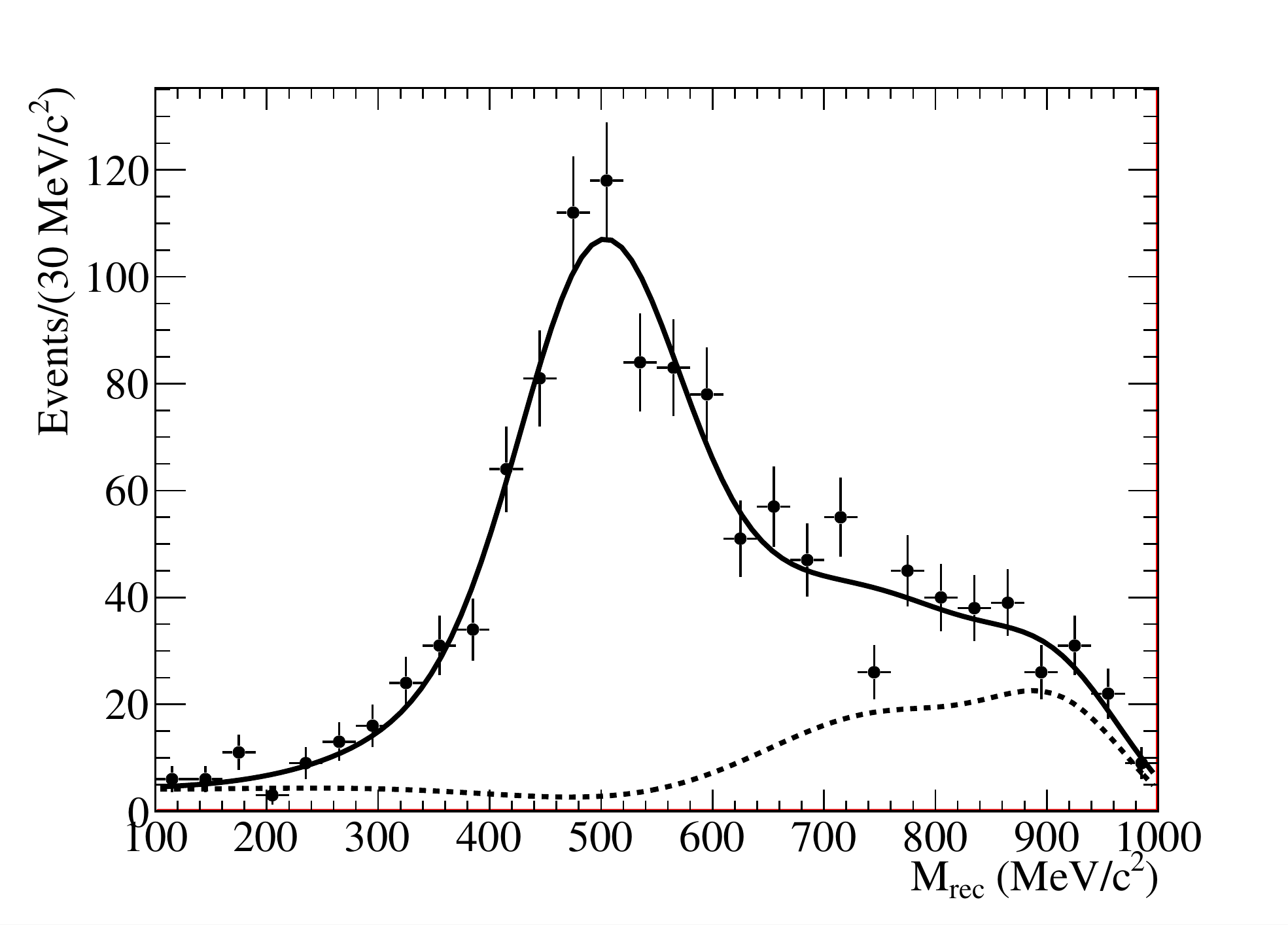}
\caption{The spectrum of the mass recoiling against the  $ K_{S} \pi^0$
system for data events from the energy region $\sqrt{s}=1.60-1.75$ GeV.
The solid curve represents the result of the fit to the spectrum with a
sum of signal and background distributions.
The dashed curve represents background distribution.
\label{fig7}}
\end{figure}

The number of signal events is determined from the fit to the $M_{\rm rec}$ 
spectrum (Fig.~\ref{fig7}) by a sum of distributions for 
signal and background events. The signal distribution is described by
a sum of three Gaussian functions with parameters determined from
the fit to the simulated signal $M_{\rm rec}$ spectrum. 
To account for a possible inaccuracy of the signal simulation, 
two parameters are introduced: mass shift $\Delta M$ and smearing parameter
$\Delta\sigma^2$. The latter is added to all Gaussian sigmas squared 
($\sigma_i^2 \to \sigma_i^2+\Delta \sigma^2$).
The parameters $\Delta M$ and $\Delta\sigma^2$ are determined from the fit
to the $M_{\rm rec}$ spectrum for events from the energy range 
$\sqrt{s}=1.6-1.75$ GeV shown in Fig.~\ref{fig7}. They are found to be
$\Delta M=(5\pm5)$ MeV/$c^2$ and $\Delta \sigma^2= 1800\pm770$ MeV$^2/c^4$. 

To obtain the background distribution we analyze simulation
for the processes
$e^+e^- \rightarrow K_{S} K_{L} $, 
$e^+e^- \rightarrow K_{S} K_{L} \pi^0 \pi^0 $,
$e^+e^- \rightarrow \phi \eta $,
$e^+e^- \rightarrow \omega \pi^0$,
$e^+e^- \rightarrow \omega \eta$,
$e^+e^- \rightarrow \omega \pi^0 \pi^0 $,
$e^+e^- \rightarrow \omega \pi^0 \eta $ with decays
$\phi \rightarrow K_{S} K_{L} $ and $\omega \rightarrow \pi^0 \gamma$.
For all these processes the existing experimental data on Born cross sections
are approximated and then used in event generators for calculation radiative
corrections, and generation of extra photons emitted from the initial state. 
The obtained simulated background distribution is fitted with a smooth function.
The largest contributions into expected background come from the processes
$e^+e^- \rightarrow K_{S} K_{L} $ and $e^+e^- \rightarrow K_{S} K_{L} \pi^0 \pi^0 $.

In the fit to the data $M_{\rm rec}$ spectra the background distribution
obtained from simulation is multiplied by a free scale factor. For
all energy regions the fitted value of the scale factor was found to be 
consistent with unity. The example of the fit for the energy region
$\sqrt{s}=1.60-1.75$ GeV is presented in Fig.~\ref{fig7}. 

Some fraction of selected signal events contains a hard photon emitted from 
the initial state. These initial-state-radiation (ISR) events have 
$M_{\rm rec}$ larger than the $K^0$ mass and distort the shape of the signal 
$M_{\rm rec}$ distribution. The distortion is most significant at energies 
$\sqrt{s}>1.92$ GeV, where the fraction of ISR events becomes larger than 50\%.
For these energies, the fitting procedure is modified. The signal
spectrum is represented as a sum of two spectra, for events with
the photon energy $E_\gamma$ smaller and larger than 120 MeV. The number of
events in the latter spectrum and its error are calculated using
the Born $e^+e^-\to K_SK_L\pi^0$ cross section measured in this work in the
lower-energy interval. The number of signal events with $E_\gamma<120$ MeV is
determined from the fit to the data $M_{\rm rec}$ spectrum.

The fitted numbers of signal events obtained for 15 energy
intervals are listed in Table~\ref{table1}.

\section{Detection efficiency\label{detef}}
The detection efficiency for $e^+e^- \rightarrow K_{S} K_{L} \pi^0$ events
is determined using MC simulation. The simulation takes into account 
radiative corrections~\cite{rad1}, in particular, emission of extra photon 
from the initial state~\cite{rad2}. The Born cross section for the process 
$e^+e^- \rightarrow K_{S} K_{L} \pi^0$ is taken from Ref.~\cite{babar}.

The detection efficiency obtained from MC simulation is corrected to take into 
account difference between data and simulation in photon conversion
in detector material before the tracking system. This difference is
measured using $e^+e^-\to \gamma\gamma$ events. The conversion probability
for two photons is canceled when normalizing to luminosity. The remaining
data-MC simulation difference for 4 photons is $-(1.35\pm0.05)\%$~\cite{ometa}.

We also study the data-simulation difference in the photon transverse 
energy-deposition  profile in the calorimeter. To do this, 
$e^+e^-\to \omega\pi^0\to\pi^0\pi^0\gamma\to 5\gamma$ events are
used, which can be selected without background~\cite{ppg_snd}. The dependence
of the photon loss due to the ``good'' photon requirement on the photon energy
is measured in data and simulation. The obtained data-simulation difference 
is used to determine the efficiency correction for simulated
$e^+e^-\to K_S K_L \pi^0$ events. The correction is found to be practically
independent of $\sqrt{s}$ and is equal to $-(3.4\pm 1.3)\%$.

The high-statistics study of the systematic uncertainty associated with 
selection of multiphoton events  based on the kinematic fit was performed
in Refs.~\cite{ppg_snd,sndompi} using $e^+e^-\to\omega\pi^0\to\pi^0\pi^0\gamma$
events. We estimate that the systematic uncertainty due
to conditions on invariant masses and $\chi^2$ of the kinematic fit does not
exceed 5\%.

The detection efficiency calculated in the model of the 
$K^\ast(892)^0 \Kbar^0$ intermediate state ($\varepsilon_{K^\ast\Kbar}$) is 
modified to take into account a small contribution of the $\phi\pi^0$ 
intermediate state:
\begin{equation}
\varepsilon=\varepsilon_{K^\ast\Kbar}(1+
\frac{\varepsilon_{\phi\pi^0}-\varepsilon_{K^\ast\Kbar}}
{\varepsilon_{K^\ast\Kbar}}f_{\phi\pi^0}),
\end{equation}
where $\varepsilon_{\phi\pi^0}$ is the detection efficiency for the process
$e^+e^-\to \phi\pi^0 \to K_SK_L\pi^0$, and $f_{\phi\pi^0}$ is the ratio of the
$e^+e^-\to \phi\pi^0 \to K_SK_L\pi^0$ cross section~\cite{babar_kkp} and
the total $e^+e^-\to K_SK_L\pi^0$ cross section obtained in this work. 
The relative difference $(\varepsilon_{\phi\pi^0}-\varepsilon_{K^\ast\Kbar})/\varepsilon_{K^\ast\Kbar}$ 
varies in the range 10--30\%. The efficiency correction is 0.1-0.7\% in the 
range 1.40--1.85 GeV, and about 2\% above and about 3\% below this interval. 
The associated systematic uncertainty is determined by the accuracy of the
$e^+e^-\to \phi\pi^0$ cross section~\cite{babar_kkp} and does not exceed
0.6\% in the range 1.4--1.9 GeV, and is 2\% above and below.
The maximum possible contribution of the $K_2^\ast(1430)^0 \Kbar^0$ mechanism
can be estimated from the measurement of the isovector and isoscalar 
$e^+e^-\to K_2^\ast(1430)^0 \Kbar^0$ cross sections in Ref.~\cite{babar_kkp}
and isospin relations~\cite{g2} assuming constructive interference of the
isovector and isoscalar amplitudes. It does not exceed is 10\% of the
total $e^+e^-\to K_SK_L\pi^0$ cross section at 
$\sqrt{s}>1.9$ GeV and negligible below. The detection efficiency for 
$e^+e^-\to K_2^\ast(1430)^0 \Kbar^0\to K_SK_L\pi^0$ events is about 40\%
larger than $\varepsilon_{K^\ast\Kbar}$. Therefore, we estimate that
the systematic uncertainty on the detection efficiency due to the
possible contribution of the $K_2^\ast(1430)^0 \Kbar^0$ mechanism 
does not exceed 4\% at $\sqrt{s}>1.9$ GeV.

The corrected detection efficiencies for the 15 energy regions are 
listed in Table~\ref{table1}. For the two intervals with $\sqrt{s}>1.92$ GeV,
the efficiency is calculated with the additional requirement $E_\gamma<120$ 
MeV. The systematic uncertainty on detection efficiency is 5.2\% in the
range $\sqrt{s}=1.4-1.9$ GeV, 5.5\% at $\sqrt{s}<1.4$ GeV, and
6.8\% at $\sqrt{s}>1.9$ GeV.

\section{The Born cross section}
The visible cross section for the process $e^+e^- \rightarrow K_{S} K_{L} \pi^0$ 
is obtained from data as:
\begin{equation}
\sigma_{{\rm vis},i}=\dfrac{N_i}{\varepsilon_i L_i}
\label{eq3}
\end{equation}
where $N_i$ is the number of $K_SK_L\pi^0$ events obtained from the fit
to the $M_{\rm rec}$ spectrum in Sec.~\ref{specfit}, $\varepsilon_i$ is the 
detection efficiency, and $L_i$ is the integral luminosity for the $i$th 
energy region.

The Born cross section $\sigma$ relates to the visible cross section as:
\begin{equation}
\sigma_{\rm vis}(\sqrt{s})=\int_{0}^{x_{\rm max}} W(s,x)\sigma(\sqrt{s(1-x)}) dx,
\label{eq4}
\end{equation}
where $ W(s,x) $ is the so-called radiator function, which describes
the probability of emission of photons with the energy $x\sqrt{s}/2$ 
by the initial electron and positron~\cite{rad1}. 

The equation (\ref{eq4}) can be represented as:
\begin{equation}
\sigma_{vis}(\sqrt{s})=\sigma(\sqrt{s})[1+\delta(\sqrt{s})],
\label{eq4a}
\end{equation}
where $\delta(\sqrt{s})$ is the radiation correction, which is calculated
as a result of the fit to the visible-cross-section data with Eq.~(\ref{eq4})
and a theoretical model for the Born cross section. The vector-meson dominance
(VMD) model~\cite{vmd} is used to describe the energy dependence of the 
$e^+e^- \rightarrow K_{S} K_{L} \pi^0$ cross section. In principle, it should 
include contributions of all vector resonances of the $\rho$, $\omega$, and 
$\phi$ families. In Ref.~\cite{babar_kkp} it is shown that the isoscalar
contribution dominates only near the maximum of the $\phi(1680)$ resonance.
Below 1.55 GeV and above 1.8 GeV the isoscalar and isovector amplitudes
are the same order of magnitude. However, for the purpose of calculating the
radiation correction, a simple model with the $\phi(1020)$ and $\phi(1680)$
resonances is sufficient. This model describes the experimental data well.
However, its fitted parameters should not be considered when measuring the 
parameters of the $\phi (1020)$ and $\phi(1680)$ resonances. 
The Born cross section for the process $e^+e^- \rightarrow K_{S} K_{L} \pi^0$
is described by the following formula:
\begin{equation}
\sigma(\sqrt{s})=\left|A_0(s)+e^{i\alpha}A_1(s)\right|^2\frac{P(s)}{s^{3/2}}
\label{eq9}
\end{equation}
where $A_0$ and $A_1$ are the amplitudes of the $\phi(1020)$ and $\phi(1680)$
decays to $K_{S}K_{L}\pi^0$, and $\alpha$ is their relative phase.
It was assumed that the decays proceed via
$K^\ast(892)^0\Kbar^0$ intermediate state. So, the function $P(s)$ describes
energy dependence of the $K^\ast(892)^0\Kbar^0$ phase space~\cite{vmd}:
\begin{eqnarray}
P(s)&=&\frac{1}{\pi}\int_{(m_{\pi^0}+m_{K^0})^2}^{(\sqrt{s}-m_{K^0})^2} 
\frac{m_{K^\ast}\Gamma_{K^\ast}}
{(q^2-m^2_{K^\ast})^2+m^2_{K^\ast}\Gamma^2_{K^\ast}}
p^3(q^2) dq^2, \label{eq5}\\
p(q^2)&=&\sqrt{\frac{(s-m^2_{K^\ast}-q^2)^2 - 4m^2_{K^\ast}q^2 }{4s}},\nonumber
\end{eqnarray}
where $m_{K^\ast}$ and $\Gamma_{K^\ast}$ are the $K^\ast(892)^0$ mass and 
width~\cite{pdg}, and $p(q^2)$ is the momentum of the $K^0\pi^0$ system.

The $\phi(1020)$ amplitude is parametrized as
\begin{equation}
A_0(s)=A_{\phi}
\frac{M_{\phi}\Gamma_{\phi}}{(M^2_{\phi}-s)-i\sqrt{s}\Gamma_{\phi}},
\label{eq7}
\end{equation}
where $A_{\phi}$ is a real constant, $M_{\phi}$ and $\Gamma_{\phi}$ are
the $\phi(1020)$ mass and width~\cite{pdg}.
while the $\phi^\prime\equiv\phi(1680)$ amplitude is given by
\begin{equation}
A_1(s)= 
\sqrt{\frac{\sigma_{\phi^\prime} M^3_{\phi^\prime}}{P(M_{\phi^\prime})}}
\frac{M_{\phi^\prime}\Gamma_{\phi^\prime}}
{(M^2_{\phi^\prime}-s)-i\sqrt{s}\Gamma_{\phi^\prime}},
\label{eq8}
\end{equation}
where $\sigma_{\phi^\prime}$ is the cross section of the process
$e^+e^- \to\phi^\prime\to K_{S} K_{L} \pi^0$ at $\sqrt{s}=M_{\phi^\prime}$,
$M_{\phi^\prime}$ and $\Gamma_{\phi^\prime}$ are the $\phi^\prime$ mass
and width. The free fit parameters are $A_{\phi}$, $\sigma_{\phi^\prime}$,
$\alpha$, $M_{\phi^\prime}$, and $\Gamma_{\phi^\prime}$. 
The model describes data well ($\chi^2/{\rm ndf}=7/15$,
where ${\rm ndf}$ is the number of degrees of freedom). The fitted
$\phi^\prime$ mass ($1700\pm 23$ MeV/$c^2$) and width ($300\pm 50$ MeV)
are close to the Particle Data Group values for $\phi(1680)$~\cite{pdg}.

\begin{table}
\caption{\label{table1}
The energy interval ($\sqrt{s}$), integrated luminosity ($L$), number of 
selected $e^+e^-\to K_SK_L\pi^0$ events ($N$), detection  efficiency 
($\varepsilon$), radiative correction factor ($1+\delta$),
and the $e^+e^-\to K_SK_L\pi^0$ Born cross section ($\sigma$).
The shown cross-section errors are statistical. The systematic error is 12\%.}
\begin{ruledtabular}
\begin{tabular}{cccccc}
$\sqrt{s}$ (GeV) &$L$ (nb$^{-1}$)&$N$&
$\varepsilon$&$1+\delta$&$\sigma$ (nb) \\
\hline
$1.300-1.350   $ & $2546$ & $18\pm 7  $& $0.058$ & 0.872 & $0.14 \pm 0.05$ \\
$1.360-1.375  $ & $1468$ & $10\pm 5  $ & $0.057$ & 0.867 &$0.14 \pm 0.06$ \\
$1.400-1.440   $ & $2783$ & $70 \pm 10$& $0.057$ & 0.851 & $0.52 \pm 0.08$ \\
$1.450-1.475  $ & $1082$ & $48\pm 9 $ & $0.057$ & 0.860 & $0.90 \pm 0.16$ \\
$1.500        $ & $2081$ & $135\pm 13$& $0.055$ & 0.867 & $1.35 \pm 0.13$ \\
$1.520-1.525  $ & $1437$ & $105\pm 12$& $0.056$ & 0.874 & $1.49 \pm 0.18$ \\
$1.550-1.575  $ & $1100$ & $135\pm 13$& $0.056$ & 0.886 & $2.48 \pm 0.24$ \\
$1.600-1.650   $ & $2997$ & $408\pm28$ & $0.054$ & 0.899 & $2.80 \pm 0.19$ \\
$1.675-1.700  $ & $2257$ & $314\pm22$ & $0.053$ & 0.923 & $2.85 \pm 0.20$ \\
$1.720-1.750   $ & $1575$ & $180\pm 17$& $0.051$ & 0.968 & $2.30 \pm 0.21$ \\
$1.760-1.800   $ & $3362$ & $311\pm 26$& $0.049$ & 1.039 & $1.81 \pm 0.15$ \\
$1.825-1.850  $ & $1973$ & $109\pm 22$& $0.049$ & 1.135 & $1.00 \pm 0.23$ \\
$1.870-1.900   $ & $3659$ & $123\pm 15$& $0.049$ & 1.249 & $0.55 \pm 0.09$ \\
$1.920-1.950   $ & $2667$ & $23\pm 7  $& $0.022$ & 0.992 & $0.39 \pm 0.12$ \\
$1.960-2.000   $ & $2458$ & $20\pm7 $  & $0.020$ & 0.974 & $0.41 \pm 0.15$ \\
\end{tabular}	  
\end{ruledtabular}
\end{table}
\begin{figure}
\centering
\includegraphics[width=0.5\textwidth]{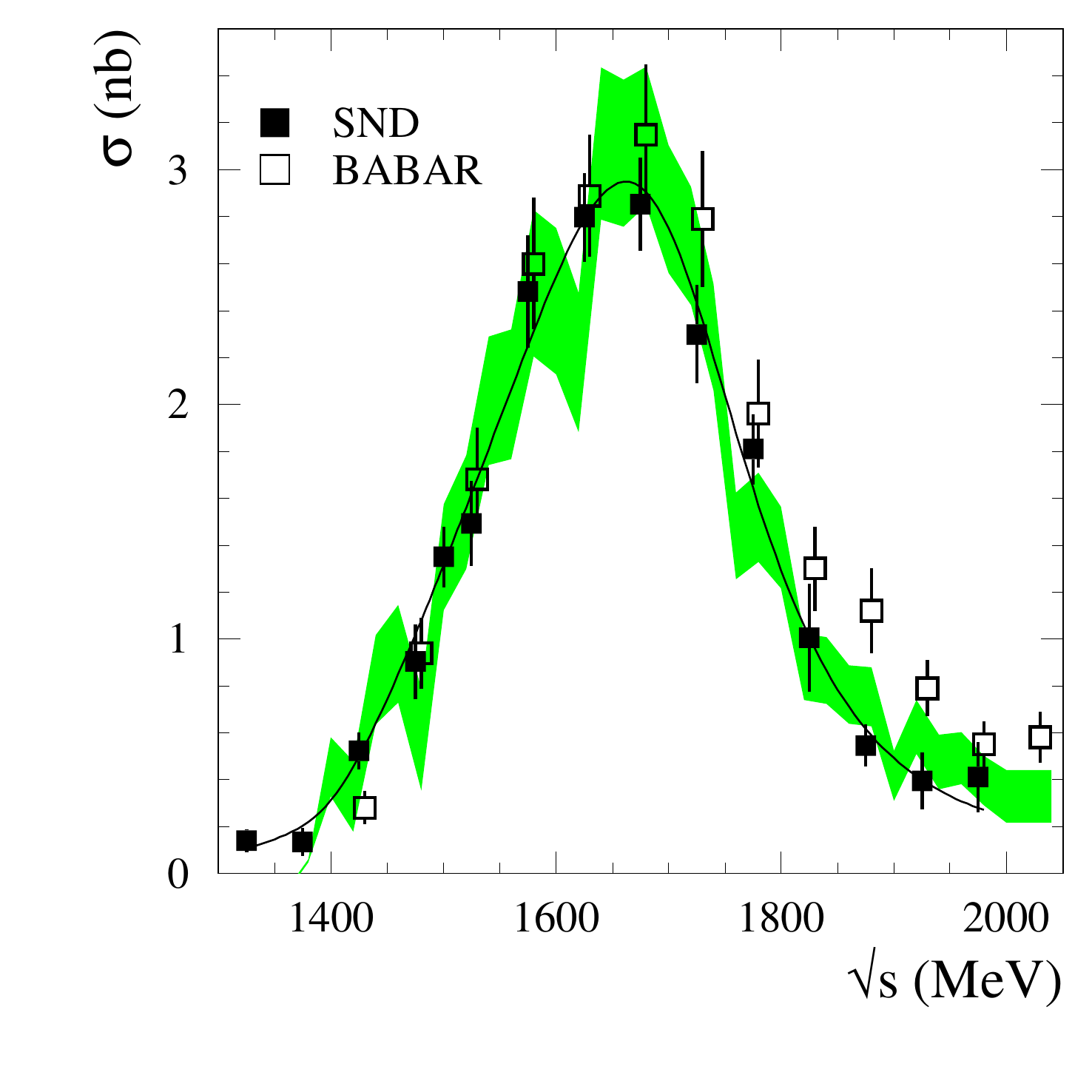}
\caption{The Born cross section for the process 
$e^+e^- \rightarrow K_{S} K_{L} \pi^0$ measured in this work 
(filled squares) in comparison with the BABAR data~\cite{babar} (open squares).
The curve represents the result of the fit to SND data with the VMD model.
The band represents the prediction for the
$e^+e^-\to K_{S} K_{L} \pi^0$ cross section
obtained using isospin relations from the BABAR measurements of
the $e^+e^-\to K_{S} K^\pm \pi^\mp$, $e^+e^-\to K^+ K^-\pi^0$, and
$e^+e^-\to\phi\pi^0$ cross sections~\cite{babar_kkp}.
\label{fig_crs}}
\end{figure}

The radiation corrections calculated with the fitted model parameters
are listed in Table~\ref{table1}. The experimental values of
the Born cross section are then obtained from the measured values of the 
visible cross sections using Eq.~(\ref{eq4a}). They are listed in 
Table~\ref{table1} and shown in Fig.~\ref{fig_crs} together with
the fitted curve.

\section{Systematic uncertainty}
Several sources give contribution to the systematic uncertainty of the 
measured cross section. These are the uncertainties of luminosity measurement
and detection efficiency, the systematic uncertainty in determination
of the number of signal events from the fit to the $M_{\rm rec}$ spectrum,
the model uncertainty of the radiative correction.

A possible source of the systematic uncertainty on the number of signal events
is imperfect simulation of the shape of the signal and background 
$M_{\rm rec}$ distributions. 

In the fit to the $M_{\rm rec}$ spectrum we use the simulated background 
distribution multiplied by a free scale factor. To estimate the systematic 
uncertainty due imperfect simulation of the background shape,
another approach to background description is applied, by a sum of predicted 
background plus a linear function. The difference in the number of signal 
events obtained with the standard and new background descriptions 
does not exceed 5\% in the energy region 1.60--1.75 GeV. This value is used 
as an estimate of the uncertainty.

The signal $M_{\rm rec}$ distribution has the asymmetric line shape
(see Fig.~\ref{fig7}). The tail of the distribution at $M_{\rm rec}>m_{K^0}$
originates from events with $N_\gamma>6$, in which wrong combination of photons
forming the $K_S\pi^0$ system is chosen. For six-photon events the line shape
is symmetric and close to Gaussian. To estimate the systematic uncertainty 
associated with the signal line shape, we repeat the analysis selecting 
events with $N_\gamma=6$. The visible cross section near the maximum is found
to be $(20\pm 5)\%$ lower than the cross section obtained with the standard 
selection criteria. The observed difference is partly explained by
incorrect simulation of $K_L$ nuclear interaction. At the $K_L$ energy
510 MeV the inelastic nuclear interaction length used in simulation~\cite{kloe}
is larger than the measured one by $(12 \pm 5)\%$. The six-photon selection in
contrast to the standard selection is very sensitive to the value of the $K_L$
nuclear interaction length: its decrease by 12\% is translated to the 11\% 
decrease of the detection efficiency~\cite{korneev}. 
The remaining difference $(9 \pm 7)\%$ is used as an 
estimate of the uncertainty associated with the signal line shape.

The model uncertainty of the radiation correction is estimated by varying the
fitted parameters of the VMD model [Eq.~(\ref{eq9})--(\ref{eq8})] within their
errors. It is below 0.5\% at $\sqrt{s}<1.65$ GeV and increases up to
3.5\% near 2 GeV. The systematic uncertainty on the detection efficiency
is discussed in Sec.~\ref{detef} and is 5.2\% in the range $\sqrt{s}=1.4-1.9$ 
GeV, 5.5\% at $\sqrt{s}<1.4$ GeV, and 6.8\% at $\sqrt{s}>1.9$ GeV.
The systematic uncertainty of the luminosity measurement studied
in Refs.~\cite{ppg_snd,sndompi} is 1.4\%.

The systematic uncertainties from different sources are summarized in 
Table~\ref{table2} for four c.m. energy intervals. The total systematic 
uncertainty including all the contributions discussed above combined in 
quadrature is estimated to be 12\% below 1.9 GeV and 13\% above.
\begin{table}
\caption{\label{table2}
The systematic uncertainties (\%) on the measured $e^+e^-\to K_SK_L\pi^0$ 
Born cross section in four c.m. energy intervals.
}
\begin{ruledtabular}
\begin{tabular}{ccccc}
Source & 1.3--1.4 GeV & 1.40--1.8 GeV & 1.8--1.9 GeV & 1.9--2.0 GeV\\
\hline
Luminosity & 1.4 & 1.4 & 1.4 & 1.4 \\
Detection efficiency & 5.5 & 5.2 & 5.2 & 6.8 \\
Background subtraction & 5.0 & 5.0 & 5.0 & 5.0\\
Signal line shape & 9.0 & 9.0 & 9.0 & 9.0\\
Radiative corrections & 0.5 & 0.5--1.2 & 1.9--2.4 & 2.6--3.5\\
\hline
Total & 12 & 12 & 12 & 13 \\
\end{tabular}
\end{ruledtabular}
\end{table}

\section{Discussion and summary}
The cross section for the process $e^+e^- \rightarrow K_{S} K_{L} \pi^0$
has been measured with the SND detector at the VEPP-2000 $e^+e^-$ collider in
the energy range 1.3--2.0 GeV. The comparison of the SND data with the only 
previous measurement, done by the BABAR Collaboration~\cite{babar},
is presented in Fig.~\ref{fig_crs}. Only statistical errors are shown. 
The systematic uncertainty of the SND data is 12--13\%, while the BABAR
systematic uncertainty increases from 10\% at 1.7 GeV and below to about 20\% 
at 2 GeV~\cite{babar}. Near the maximum of the cross section (1.7 GeV) 
the SND points lie below the BABAR points, but agree within systematic errors.
The same trend persists at higher energies, up to 2 GeV. 
The largest difference, about 2 standard deviations including systematic
uncertainties, between the SND and BABAR data is observed in the energy 
points 1.875 and 1.925 GeV.

It is discussed in Sec.~\ref{intst} that the dominant mechanism of the 
$e^+e^- \rightarrow K_{S} K_{L} \pi^0$ reaction at $\sqrt{s} < 2$ GeV
is $K^\ast(892)^0 \Kbar^0$. Under this assumption the 
cross section of the process under study can be predicted using
the isospin relation~\cite{g2}
\begin{eqnarray}
\sigma(e^+e^-\to K_{S} K_{L} \pi^0)&=&
\sigma(e^+e^-\to K_{S} K^\pm \pi^\mp)-
\sigma(e^+e^-\to K^+ K^-\pi^0)+\label{isospin}\\
& & B(\phi\to K\Kbar)\sigma(e^+e^-\to\phi\pi^0)\nonumber
\end{eqnarray}
and the BABAR measurements~\cite{babar_kkp} of the 
$e^+e^-\to K_{S} K^\pm \pi^\mp$, $e^+e^-\to K^+ K^-\pi^0$, and
$e^+e^-\to\phi\pi^0$ cross sections. In Eq.~(\ref{isospin}) we take
into account that both $\sigma(e^+e^-\to K_{S} K_{L} \pi^0)$
and $\sigma(e^+e^-\to K^+ K^-\pi^0)$ contain contributions
of the $\phi\pi^0$ intermediate state.
The predicted cross section is shown in Fig.~\ref{fig_crs} 
by the green band and is found to be in good agreement with our measurement.

\section{ACKNOWLEDGMENTS}
This work is supported by the RFBR grants 16-02-00014 and 16-02-00327.
Part of this work related to the photon reconstruction algorithm in the
electromagnetic calorimeter for multiphoton events is supported
by the Russian Science Foundation (project No. 14-50-00080).

\end{document}